\documentstyle[preprint,aps,prl]{revtex}
\begin{document}
\bibliographystyle{prsty}
\draft 		
\title{Theory of Electronic Raman Scattering in Disordered $d-$wave
Superconductors}
\author{T. P. Devereaux}
\address{Department of Physics, University of California, Davis, CA 95616}
\maketitle
\date{\today}
\begin{abstract}
A theory for the effect of impurity scattering on electronic Raman
scattering in unconventional superconductors is presented. The impurity
dependence of the the spectra can be used to to distinguish between
conventional (anisotropic $s-$wave) or unconventional ($d$-wave) energy
gaps, and excellent agreement with the data on three cuprate
superconductors is shown for $d_{x^{2}-y^{2}}$ pairing. The full
frequency dependence of the spectra can be reproduced by including
inelastic scattering due to spin fluctuations and/or Coulomb scattering
on a partially nested Fermi surface.
\end{abstract}
\narrowtext

By now a more coherent picture of the pair state symmetry in
cuprate superconductors has
emerged, and it has become increasingly evident
that a gap with {\it predominantly} $d_{x^{2}-y^{2}}$ pairing symmetry
can provide an adequate description of many experiments\cite{pines}.
However, the
question remains or not whether the order parameter changes sign around the
Fermi surface.
The available data on Josephson
junction and point-contact tunneling measurements seems to in favor
of the former, although examples exist to the contrary
\cite{jj,pines}.
In addition, it has been argued that the impurity dependence of
certain correlation functions can be used to determine whether the
gap changes sign.  In particular, a detailed analysis has been given
for the case of the electromagnetic penetration depth\cite{pjh}, and
the crossover from $T$ to $T^{2}$ at a temperature $T^{*}$ which
grows with impurity concentration has be taken to support $d-$wave
pairing. Similar considerations can be made for the Knight shift,
NMR rates, etc. \cite{pines}. However, the evidence gained from
the penetration depth or other correlation functions is incomplete in that
only the topology of the energy gap nodes can be identified
from the power-law behavior at low temperatures, which leaves the actual gap
symmetry unspecified. For instance, in tetragonal
superconductors, there are 5 pure representations, ($d_{x^{2}-y^{2}}, d_{xy},
d_{3r^{2}-1}, d_{xz}$ and $d_{yz}$), which yield
the same low temperature power-laws and impurity dependences for, e.g.,
the penetration depth, NMR rates, density of states, etc.

It has recently been shown that both the electronic and
phonon contributions to Raman scattering in superconductors can be
used to identify the order parameter symmetry and can distinguish
between the various representations for $d-$wave
pairing\cite{tpd}. Raman scattering measures effective intracell density
fluctuations $\tilde\rho_{\bf q}$,
\begin{equation}
\tilde\rho_{\bf q}=\sum_{{\bf k},\sigma}\gamma({\bf k})
c^{\dagger}_{{\bf k-q/2},\sigma} c_{{\bf k+q/2},\sigma},
\end{equation}
where $c^{\dagger}_{{\bf k},\sigma} (c_{{\bf k},\sigma})$ is the
creation (annihilation) operator for an electron with wavevector ${\bf k}$
and spin $\sigma$. For the case of non-resonant scattering\cite{multi},
the Raman vertex $\gamma$ is directly related to the
curvature of the band dispersion $\epsilon({\bf k})$
and incident ${\bf e^{I}}$ and scattered
${\bf e^{S}}$ polarization light vectors via
\begin{equation}
\gamma({\bf k})={m\over{\hbar^{2}}}\sum_{\mu,\nu}e^{I}_{\mu}
{\partial^{2}\epsilon({\bf k})\over{\partial k_{\mu} \partial k_{\nu}}}
e^{S}_{\nu},
\end{equation}
which can be expanded in terms of a complete set of orthogonal functions
on the Fermi surface, $\gamma({\bf k})=\sum_{L} \gamma_{L} \Phi_{L}({\bf k})$.
For bands with non-parabolic dispersion, all $L$ symmetry
channels can in principle contribute to the vertex.
Thereby, fluctuations on different parts of the Fermi surface
can be selected by orienting the polarization light vectors.

It was shown in Ref. \cite{tpd} that the
symmetry- (polarization-) dependence of the Raman spectra
straightforwardly arises through the coupling of the Raman
vertex to an anisotropic energy gap.
The main features of
the theory for clean superconductors with $d_{x^{2}-y^{2}}$
pairing symmetry are as follows:
1) due to the presence of the nodes,
no clear gap threshold is seen for any polarization channel, 2) the peak in the
spectra occurs at the largest frequency for
polarizations which select the $B_{1g}$ channel (i.e, for YBCO, crossed
in-lane polarizations aligned 45 degrees with respect to the $a-b$ axes),
while lower peaks are
observed in the $B_{2g}$ (45 degree rotation of the $B_{1g}$ orientation)
and $A_{1g}$ (mostly parallel polarizations) channels in descending order,
respectively, and 3) the low frequency behavior is given by
$\omega^{3}$ for the $B_{1g}$ channel, while is linear in $\omega$ for
the other channels. Such a strong polarization dependence of the
spectra allowed for a stringent test of the predictions for
$d_{x^{2}-y^{2}}$ pairing, and good agreement of the theory was shown
with the data on three cuprate materials, YBa$_{2}$Cu$_{3}$O$_{7}$,
Bi$_{2}$Sr$_{2}$CaCu$_{2}$O$_{8}$, and Tl$_{2}$Ba$_{2}$CuO$_{6}$.

However, like other correlation functions,
only $\mid\Delta({\bf k})\mid$ could be
determined from the fit, and thus it could not be ascertained whether
the gap changed sign around the Fermi surface. In addition, the fit was
less satisfactory for samples with
increasing disorder, with the best agreement found for a sample of
Bi 2:2:1:2 with a transition temperature $\sim 86$ K.
The theory also failed to predict {\it any} Raman
intensity in the normal state due
to phase space restrictions (consequence of the limit ${\bf q}\rightarrow 0$),
and a large and unspecified smearing width
$\Gamma \sim 0.2\Delta_{0}$ was needed even at low temperatures.

The purpose of this Letter is to show that impurity scattering can
resolve the above deficiencies of the theory.
The $T-$matrix approach to incorporate repeated scattering events
from a single impurity site has been well documented\cite{hew}.
The two parameters to
describe the theory are the cotangent of the scattering phase shift,
$c=\cot (\delta)$, and the impurity concentration $n_{i}$
described through the scattering rate $\Gamma=n_{i}/\pi N_{F}$, where
$N_{F}$ is the density of states per spin at the Fermi level.  To
maintain gauge invariance, the effects of impurity scattering are
included into both the self energy and vertex renormalizations. The
case for an even parity vertex (such as the Raman vertex) has been worked
out for general pair-states under the assumption of particle-hole symmetry and
s-wave impurity scattering in Ref.
\cite{hew}, and thus we can adopt these methods accordingly.
For the case of the Raman vertex, impurity dressings
are only important for the case of the density ($L=0$) channel, while other
channels do not couple to the scattering. Since the density
channels are completely screened out (intercell charge fluctuations)
for vanishing momentum transfers, for our purposes we
can set the vertex corrections to zero. This will not hold for
an impurity interaction which is non-zero in other channels (i.e.,
a {\bf k}-dependent impurity interaction). This can be remedied
by including couplings to other channels, but
the analysis can not be carried as far analytically.

Our expression for the Raman response is then given by
\begin{equation}
\chi^{\prime\prime}({\bf q}\rightarrow 0,i\Omega)=
-T\sum_{i\omega} \sum_{\bf k}Tr \gamma({\bf k}) \hat \tau_{3}
\hat G(k, i\omega) \hat\tau_{3}\gamma({\bf k})
\hat G({\bf k},i\omega-i\Omega),
\end{equation}
where $Tr$ denotes the trace, $\hat\tau$ are Pauli matrices,
and $\gamma$ is the bare Raman vertex, Eq. (2).
We use the standard weak coupling BCS Green's function
dressed by the impurity self energy,
\begin{equation}
\hat G({\bf k},i\omega)={i\tilde\omega+\tilde\epsilon({\bf k})\hat\tau_{3}
+\Delta({\bf k})\hat\tau_{1}
\over{(i\tilde\omega)^{2}-\tilde\epsilon({\bf k})^{2}-\Delta({\bf k})^{2}}},
\end{equation}
where the tilde indicates the renormalized frequency and band energy via
\begin{eqnarray}
i\tilde\omega=i\omega-\Sigma_{0}(i\tilde\omega), \nonumber \\
\tilde\epsilon({\bf k})=\epsilon({\bf k})-\Sigma_{3}(i\tilde\omega).
\end{eqnarray}
Lastly, the matrix self energy is given in terms of the integrated
Green's function $g(i\omega)={1\over{\pi N_{F}}}\sum_{{\bf k}}Tr \hat\tau_{0}
\hat G({\bf k}, i\omega)$,
\begin{equation}
\hat\Sigma(i\omega)=\Gamma{g(i\omega)\hat\tau_{0}-c\hat\tau_{3}\over{
c^{2}-g^{2}(i\omega)}}=\Sigma_{0}\hat\tau_{0}+\Sigma_{3}\hat\tau_{3}.
\end{equation}
Our choice of the gap parameter is $\Delta({\bf k})=\Delta_{0}\cos(2\phi)$,
where we will for simplicity chose to work with a 2-D (cylindrical)
Fermi surface.

The results of the theory with the inclusion of impurity scattering
in the Born ($c >> 1$) and unitary ($c=0$) limits are shown in
Figure 1. The impurity scattering smears out the sharp
features of the spectra (i.e., the logarithmic divergence of the
spectra at the gap edge for the $B_{1g}$ channel) and thus can
provide an explanation for the broadening seen in the data at
low temperatures. In addition, the anisotropy of the peak
positions is maintained (unless the scattering is so large that
the normal state behavior is recovered). The coupling of the gap and Raman
vertex lead to polarization dependent peak positions, with $\omega_{peak}/
\Delta_{0}=2.0, 1.8,$ and $1.2$ for the $B_{1g}, B_{2g},$ and $A_{1g}$
channels, respectively. Thereby, the channel which shows the peak at the
highest frequency gives the predominant symmetry of the energy gap.  This
was used in Ref. \cite{tpd} to conclude that the energy gap in the cuprates
is {\it predominantly} of $d_{x^{2}-y^{2}}$ symmetry as opposed to $d_{xy}$
or other $d-$wave representations.

To determine further whether the gap
is {\it entirely} of $d_{x^{2}-y^{2}}$ symmetry, the impurity dependence of
the low frequency behavior of the channel dependent spectra can be exploited.
{}From Figure 1, it is seen that Born impurity scattering yields low
frequency exponents that are the same as for clean materials
(unless of course $\Gamma \sim \Delta_{0}$ and the normal state is recovered).
However, in the unitary case, while the low frequency behavior remains
linear in
frequency in the $A_{1g}$ and $B_{2g}$ channels, below a
characteristic frequency $\omega^{*}$ the behavior changes from
$\omega^{3}$ to linear in $\omega$ for the $B_{1g}$ channel.
This is due to a nonzero density of states at the Fermi level,
which allows for normal-state-like behavior to be recovered \cite{gorkov}.
As
in the case of the penetration depth\cite{pjh}, the scale
$\omega^{*}$ grows with increasing impurity concentrations.
However, the exponent is
symmetry dependent (remains 1 for $A_{1g}$ and $B_{2g}$ channels,
while {\it decreases} from 3 to 1 for the $B_{1g}$ channel, which is
unlike the penetration depth).
This is in marked contrast to the impurity dependence of the
spectra for anisotropic $s$-wave gaps, since in that case, the
exponents for {\it all} channels would grow as the gap becomes
more isotropic for increased impurity scattering \cite{norman}.
In particular, the impurity dependence of the $B_{1g}$ channel
is opposite to what one would expect if the gap was anisotropic
$s$-wave. Thus the
impurity dependence can be systematically checked to determine
whether the gap has accidental or intrinsic zeroes, or more
generally, if the gap is anisotropic $s$-wave with predominantly
$B_{1g}$ symmetry, or if it is $d_{x^{2}-y^{2}}$ symmetry.

To make a comparison to experiment, we note that the impurity
dependence, for instance via Zn doping,
has not been systematically checked via Raman scattering
for any of the cuprate superconductors. Raman scattering
measurements of the electronic background have been performed on
materials of varying quality. In particular, we focus on three materials,
as-grown Bi$_{2}$Sr$_{2}$CaCu$_{2}$O$_{8}\ ({\rm T}_{c}=86 K)$,
oxygen-annealed Bi$_{2}$Sr$_{2}$CaCu$_{2}$O$_{8}\ ({\rm T}_{c}=79 K)$
\cite{hackl1},
and Tl$_{2}$Ba$_{2}$CuO$_{6}\ ({\rm T}_{c}=80 K)$ \cite{hackl2}.
The materials show polarization dependent
spectra, with the peak positions accurately predicted using a gap of
$d_{x^{2}-y^{2}}$ symmetry (see Ref. \cite{tpd}). These
compounds differ with respect to both surface quality and the presence
of bulk impurity defects.
Fits using the
theory for clean materials\cite{tpd} required a Gaussian broadening width
$\Gamma \sim 0.2 \Delta_{0}$.
Such large broadening, if attributed solely
to impurity scattering, would of course also lead to a large reduction
of $T_{c}$. However, the majority of the smearing could be due to a
distribution of T$_{c}$ values (of the order of 20 percent in Tl
2:2:0:1) and inelastic
quasiparticle scattering, which could be taken into account via strong
coupling theory.  The actual magnitude of the scattering to be
attributed to impurities remains an open issue which requires both a
non-phenomenological
theory to include the effects of inelastic scattering, doping,
and disorder on the pairing interactions, as well as a systematic
experimental analysis.

We present the comparison of the theory to the spectra obtained in
Refs. \cite{hackl1,hackl2} on the three compounds in Figure 2.
The spectra shown in the Figure are for the $B_{1g}$ and $B_{2g}$
channels for both Bi 2:2:1:2 samples, and for $B_{1g}$ and $A_{1g}$ for
Tl 2:2:0:1.
The parameters used to fit the data are
$\Delta_{0}=
287,\ 192,$ and $240\ {\rm cm}^{-1}$, for the energy gap, and resonant
impurity scattering rate
$\Gamma/\Delta_{0}=0.125,\ 0.2,$ and $0.25$ for as-grown, oxygen-
annealed Bi 2:2:1:2,  and Tl 2:2:0:1, respectively. No Gaussian broadening was
used.
The theory shows that the anisotropy
of the peak positions is preserved even in the presence of impurity
scattering which suggests that the gap anisotropy is robust to disorder.
The linear rise
in frequency of {\it both} the $B_{1g}$ and $A_{1g}$ channels is a
direct consequence of resonant impurity scattering.
This is shown in log-log plot of the low
frequency part of the $B_{1g}$ response for the three materials in the inset of
Figure 2. Born scattering
can not fit the data as the amount of scattering to produce the linear
rise in the $B_{1g}$ channel fully suppresses the peak feature.
The crossover scale
$\omega^{*}$ grows with increasing disorder, which leads to a more
pronounced linear rise of
the $B_{1g}$ channel.
The crossover frequency is given by $\omega^{*}/\Delta_{0}=
0.38, 0.47,$ and $0.73$ for the as-grown, oxygen-annealed Bi 2:2:1:2, and
Tl 2:2:0:1,
respectively. This is in consistent with the predictions of $d-$wave
pairing and is difficult to reconcile using an anisotropic $s-$wave
gap. Thus, this crucial issue merits a systematic experimental
examination of the low frequency region of the spectra using greater
resolution.

The theoretically predicted spectra fail to reproduce the
flat background extending out to large Raman shifts\cite{normal} since
the calculated spectra falls off as $1/\omega$\cite{zawa}.
``Marginal-like'' behavior
\cite{marg}
can be phenomenologically modelled by inserting a self energy piece of
the form
\begin{equation}
\Sigma^{\prime\prime}(\omega)=\alpha \sqrt{\omega^{2}+\beta T^{2}},
\end{equation}
where $\alpha$ and $\beta$ are constants. This form for the self-energy
arises in the normal state when electron-electron scattering
on a nested Fermi surface is
taken into account\cite{nest} or spin-fluctuation scattering is considered
\cite{NFL}.  While the inclusion of this term to
the self energy does not take the modifications of superconductivity
into account and also neglects vertex corrections of the interactions, the
flat behavior at high frequencies can be reproduced to fit the data.
This is shown explicitly in Figure
3, which compares the results of the theory with and without
the added self energy to the $B_{1g}$ data on the annealed Bi 2:2:1:2 sample
using a value of the gap, $\Delta_{0}=190\ {\rm cm}^{-1}$. The parameters
used are $\alpha/\Delta_{0}=0.25, \beta=3.3$, and small
resonant scattering $\Gamma=0.01\Delta_{0}$ for the solid line,
which is to be compared to
the fit made in Figure 2 without the self energy term added (redrawn in
Figure 3 as the dotted line).
While the low frequency portion of the spectrum $\omega < 2\Delta_{0}$
is relatively unaffected, the higher frequency region is well fit by
the self energy inclusion. It is expected that a better fit could be
obtained when modifications due to superconductivity are included in
the interaction self energy.

The agreement of the theory to the data leads to the following conclusions:
First, the anisotropy of the peak positions result from a coupling of
the Raman vertex to an energy gap with ${\it predominantly}$ $d_{x^{2}-y^{2}}$
symmetry. Second, since the spectra do not show
evidence for a reorganization of states to higher frequencies with increasing
disorder, and on the contrary, reorganize to lower frequencies, this
strongly suggests that
the gap therefore changes sign and has unconventional
$d_{x^{2}-y^{2}}$ symmetry as opposed to conventional anisotropic $s$-wave
symmetry. This is seen from the increased prominence of the linear rise
of the $B_{1g}$ spectra and the increase of $\omega^{*}$
for increasing resonant scattering.
Lastly, the full frequency range of the spectra
can be modelled by taking into account an additional
interaction self energy.  A systematic check of
impurity scattering on a single compound would clarify these points
further.

The author would like to thank D. Einzel, R. Hackl, P. Hirschfeld,
D. Pines, R. R. P. Singh, D. Scalapino, R. T. Scalettar, A.
Virosztek, A. Zawadowski, and G. Zimanyi for many enlightening discussions.
This work was supported by NSF Grant Number 92-06023.

\begin{figure}
\caption{Raman spectra for $B_{1g}, B_{2g},$ and $A_{1g}$ channels, with
values of $c$ and $\Gamma$ indicated. Magnitude of the Raman vertices are set
to 1. The inset shows the low frequency behavior of
the $B_{1g}$ channel. For the case of resonant scatterers ($c=0$, bottom
panel), the low frequency behavior crosses over from $\omega$ to
$\omega^{3}$.}
\end{figure}

\begin{figure}
\caption{Fit to the data taken for the $B_{1g}$ and $B_{2g}$ channels in
as-grown Bi 2:2:1:2 ($\rm{T}_{c}=86 K$, top panel), O$_{2}$ annealed Bi
2:2:1:2
($\rm{T}_{c}=79 K$, middle panel) and for the $B_{1g}$ and $A_{1g}$
channels in Tl 2:2:0:1 ($\rm{T}_{c}= 80 K $, bottom panel) obtained in
Refs. \protect\onlinecite{hackl1} and \protect\onlinecite{hackl2}.
The phonon contributions have been
subtracted out in the middle and bottom panels. The parameters used
are $\Delta_{0}=287, 190,$and $240\ {\rm cm}^{-1},\ \Gamma/\Delta_{0}=0.125,\
0.2,$ and $0.25$, for the top, middle, and bottom panels, respectively.
Inset: Log-log plot of the low frequency portion of the $B_{1g}$ response.
The crossover frequency $\omega^{*}/\Delta_{0}=0.38, 0.47,$ and $0.73$ for
the top, middle, and bottom panels, respectively.}
\end{figure}

\begin{figure}
\caption{Theory curves calculated with and without interaction self energy,
Eq. 7, compared to the Raman spectrum for the $B_{1g}$ channel of the
O$_{2}$- annealed Bi 2:2:1:2 sample. The dotted line is redrawn from Fig. 2.}
\end{figure}

\end{document}